\documentclass[aps,prb,amsmath,amssymb,reprint]{revtex4-1}
\usepackage{units}
\usepackage{graphicx}

\begin{document}

\title{Magneto-optical imaging of voltage-controlled magnetization reorientation}

\author{A.~Brandlmaier}
\email{andreas.brandlmaier@wmi.badw.de}
\affiliation{Walther-Mei{\ss}ner-Institut, Bayerische Akademie der Wissenschaften, 85748 Garching, Germany}

\author{M.~Brasse}
\affiliation{Walther-Mei{\ss}ner-Institut, Bayerische Akademie der Wissenschaften, 85748 Garching, Germany}

\author{S.~Gepr\"{a}gs}
\affiliation{Walther-Mei{\ss}ner-Institut, Bayerische Akademie der Wissenschaften, 85748 Garching, Germany}

\author{M.~Weiler}
\affiliation{Walther-Mei{\ss}ner-Institut, Bayerische Akademie der Wissenschaften, 85748 Garching, Germany}

\author{R.~Gross}
\affiliation{Walther-Mei{\ss}ner-Institut, Bayerische Akademie der Wissenschaften, 85748 Garching, Germany}
\affiliation{Physik-Department E23, Technische Universit\"{a}t M\"{u}nchen, 85748 Garching, Germany}

\author{S.~T.~B.~Goennenwein}
\affiliation{Walther-Mei{\ss}ner-Institut, Bayerische Akademie der Wissenschaften, 85748 Garching, Germany}
\affiliation{Physik-Department E23, Technische Universit\"{a}t M\"{u}nchen, 85748 Garching, Germany}

\date{\today}

\begin{abstract}
We study the validity and limitations of a macrospin model to describe
the voltage-controlled manipulation of ferromagnetic magnetization
in nickel thin film/piezoelectric actuator hybrid structures. To this
end, we correlate simultaneously measured spatially resolved magneto-optical
Kerr effect imaging and integral magnetotransport measurements at
room temperature. Our results show that a macrospin approach is adequate
to model the magnetoresistance as a function of the voltage applied
to the hybrid, except for a narrow region around the
coercive field---where the magnetization reorientation evolves via
domain effects. Thus, on length scales much larger than the typical
magnetic domain size, the voltage control of magnetization is well
reproduced by a simple Stoner-Wohlfarth type macrospin model.
\end{abstract}

\maketitle

\section{Introduction}

Multifunctional material systems are of great technological interest
\cite{Ramesh:NatMater:6:2007}. Amongst this class, magnetoelectric
multiferroics are of particular relevance, as they enable an \emph{in-situ}
electric-field control of magnetization \cite{Fiebig:JPhysD:38:2005,Eerenstein:Nature:442:2006,Zhao:NatMater:5:2006}.
Regarding device applications, composite-type multifunctional structures
constitute an appealing approach \cite{Zavaliche:NanoLett:5:2005,Chu:NatMater:7:2008,Wu:NatMater:9:2010,Bibes:NatMater:7:2008,Binek:JPhys:17:2005,Stolichnov:NatMater:7:2008,Mathews:Science:276:1997},
and thus are extensively investigated. Since a local magnetization control
is of particular interest, an imaging of the spatial evolution of $\mathbf{M}$ due
to strain-mediated magnetoelectric coupling is mandatory
\cite{Zavaliche:NanoLett:5:2005,Brintlinger:NanoLett:10:2010,Taniyama:JAP:101:2007,Chung:APL:92:2008,Chung:APL:94:2009,Xie:JAP:108:2010}. However, most reports on magnetization
changes $M\left(E\right)$ as a function of the applied electric field
rely on either integral measurement techniques or magnetic force microscopy
imaging \cite{Taniyama:JAP:101:2007,Chung:APL:92:2008,Chung:APL:94:2009,Xie:JAP:108:2010}.
In contrast, spatially resolved experiments to address local $M\left(E\right)$
changes on macroscopic ($\unit{mm^{2}}$) areas are scarce.
We here focus on ferromagnetic/ferroelectric hybrid systems, in which
a strain-mediated, indirect magnetoelectric coupling via the magnetoelastic
effect is exploited \cite{Eerenstein:NatMater:6:2007,Thiele:PRB:75:2007,Nan:JAP:103:2008,Israel:NatMater:7:2008,Sahoo:PRB:76:2007,Zheng:Science:303:2004,Gepraegs:APL:96:2010,Liu:AdvFunctMater:19:2009,Chen:APL:97:2010,Kim:JMMM:267:2003,Boukari:JAP:101:2007,Chen:APL:94:2009}.
More specifically, we study multifunctional hybrids \cite{Gepraegs:PhilosMagLett:87:2007}
composed of ferromagnetic nickel thin films and bulk piezoelectric
actuators as {}``spin-mechanics'' model systems, in which a voltage-controlled
strain is induced in the ferromagnets \cite{Brandlmaier:PRB:77:2008,Bihler:PRB:78:2008,Goennenwein:PSS:2:2008,Weiler:NJP:11:2009}.
Using spatially resolved magneto-optical Kerr effect imaging, we investigate
the magnetization evolution in our samples both as a function of external
magnetic field and of electrical voltage applied to the actuator.
We observe that the magnetization mainly reorients by coherent and
continuous rotation. Only for a small region around the coercive field
the magnetization reorientation proceeds via domain formation and
propagation. To quantitatively evaluate the Kerr images, we extract
a macrospin corresponding to an effective, average magnetization orientation
by spatially averaging over regions of interest in the images. Comparing
the anisotropic magnetoresistance calculated using this macrospin
with the corresponding measurements yields excellent agreement. This
corroborates the notion that a macrospin picture is adequate except
for a narrow region around the coercive field. Having established
the applicability of a macrospin approach, we quantitatively evaluate
the voltage-controlled changes of the magnetization orientation and
the reversibility of the voltage-induced magnetization reorientation
using magnetotransport techniques.

\section{Experiment}

\begin{figure}
\begin{centering}
\includegraphics[width=1\columnwidth]{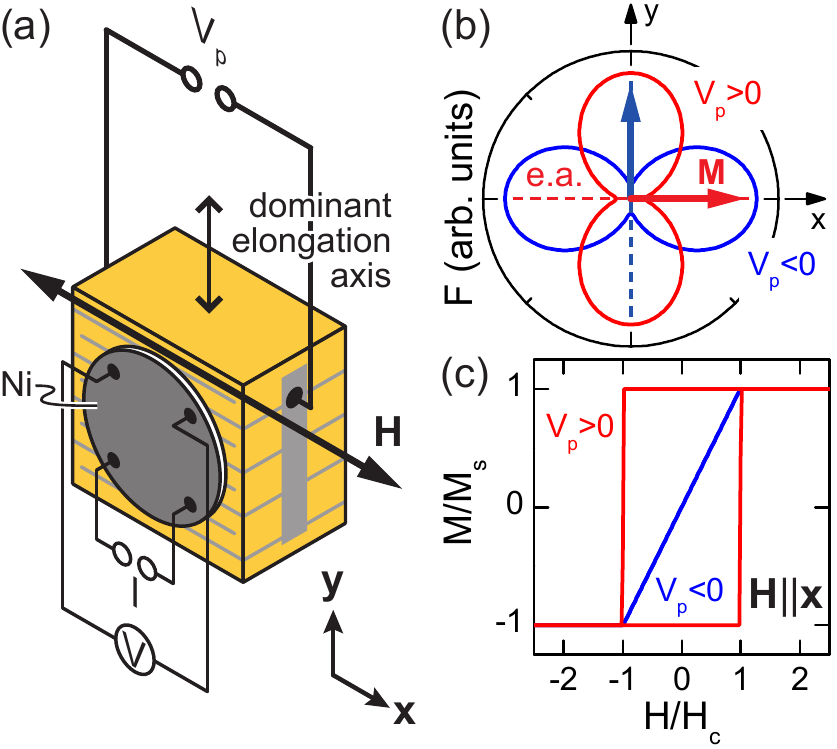}
\par\end{centering}

\caption{(a) Schematic illustration of the ferromagnetic thin film/actuator
hybrid, indicating the orientation of the external magnetic field
$\mathbf{H}\parallel\mathbf{x}$ and the contact scheme in four-point
geometry for AMR measurements. (b) Free energy density contours in
the $\mathbf{x}-\mathbf{y}$ film plane for different $V_{\mathrm{p}}$
applied, with the orientation of the magnetic easy axis (e.a.) depicted
by the dashed lines. Thus, the equilibrium magnetization orientation
is oriented along $\mathbf{x}$ for $V_{\mathrm{p}}>\unit[0]{V}$
(red line), while it is aligned along $\mathbf{y}$ for $V_{\mathrm{p}}<\unit[0]{V}$
(blue line). (c) Calculated $M\left(H\right)=\left(\mathbf{M}\cdot\mathbf{x}\right)/M_{\mathrm{s}}$
loops with $\mathbf{H}\parallel\mathbf{x}$ using a single-domain
Stoner-Wohlfarth model, as explained in the text. In accordance to
(b), the $M\left(H\right)$ curves show a magnetically easy $\mathbf{x}$
direction by the rectangular-shaped loop for $V_{\mathrm{p}}>\unit[0]{V}$
(red line), while the non-hysteretic line indicates a magnetically
hard direction for $V_{\mathrm{p}}<\unit[0]{V}$ (blue line).\label{fig:concept}}
\end{figure}
We realize a voltage control of magnetization in ferromagnetic thin
film/piezoelectric actuator hybrid structures. The commercially available,
co-fired $\mathrm{Pb}\left(\mathrm{Zr}_{x}\mathrm{Ti}_{1-x}\right)\mathrm{O}_{3}$
(PZT) piezoelectric multilayer stack actuators \textquotedblleft{}PSt
$150/2\times3/5$\textquotedblright{} (Piezomechanik M\"{u}nchen) {[}Fig.~\ref{fig:concept}(a){]}
comprise an interdigitated approx.\ $\unit[100]{\mu m}$ thick PZT-ceramic/AgPd
electrode layer structure. For the ferromagnet, we use polycrystalline
nickel (Ni) as a generic itinerant 3d magnet with a Curie temperature
$T_{\mathrm{C}}>\unit[600]{K}$ (Ref.~\onlinecite{Leger:PRB:6:1972}) well
above room temperature, a high bulk saturation magnetization $M_{\mathrm{s}}=\unit[493]{kA/m}$
(Ref.~\onlinecite{Pauthenet:JAP:53:1982}), a considerable polycrystalline
volume magnetostriction $\bar{\lambda}=\frac{2}{5}\lambda_{100}+\frac{3}{5}\lambda_{111}=-32.9\times10^{-6}$
(Ref.~\onlinecite{Lee:RPP:18:1955}), and a moderate anisotropic magnetoresistance
(AMR) ratio $\Delta\rho/\rho_{0}=\unit[2]{\%}$ (Ref.~\onlinecite{McGuire:IEEETransMagn:11:1975}).
To allow for an optimized interfacial strain coupling between the
assembled ferromagnetic and piezoelectric compounds, $\unit[100]{nm}$
thick Ni films were directly deposited onto an area of $\unit[3]{mm^{2}}$
on the actuators by electron beam evaporation at a base pressure of
$\unit[7.0\times10^{-8}]{mbar}$. The Ni films were covered in situ
by $\unit[5]{nm}$ thick Au films to prevent oxidation. Prior to the
evaporation process, a $\unit[140]{nm}$ thick polymethylmethacrylate
(PMMA) layer was spin-coated onto the respective actuator face and
baked at $\unit[110]{^{\circ}C}$ to electrically isolate the Ni film
from the actuator electrodes.

As schematically sketched in Fig.~\ref{fig:concept}(a), the actuator
deforms upon the application of a voltage $V_{\mathrm{p}}\neq\unit[0]{V}$,
exhibiting a dominant elongation axis along $\mathbf{y}$ with a maximum
strain $\epsilon_{y}=1.3\times10^{-3}$ in the full voltage swing
$\unit[-30]{V}\leq V_{\mathrm{p}}\leq\unit[+150]{V}$ (Ref.~\onlinecite{piezomechanik:booklet:LowVoltStacks:2010}).
Thus the ferromagnetic/piezoelectric hybrid allows to induce an electrically
tunable strain in the Ni thin film. In particular, a voltage $V_{\mathrm{p}}>\unit[0]{V}$
($V_{\mathrm{p}}<\unit[0]{V}$) results in an elongation with a related
strain $\epsilon_{y}>0$ (contraction with $\epsilon_{y}<0$) along
$\mathbf{y}$. Due to elasticity, such a tensile (compressive) strain
along $\mathbf{y}$ is accompanied by a contraction (elongation) along
the orthogonal in-plane direction $\mathbf{x}$, with about half the
magnitude.

To quantify the effect of lateral elastic stress on the magnetic anisotropy
of a ferromagnetic thin film, we rely on a magnetic free-energy model.
Since this approach is discussed in detail, e.g., in Ref.~\onlinecite{Weiler:NJP:11:2009},
we here only qualitatively summarize our findings. Because we use polycrystalline
ferromagnetic thin films, which exhibit no net crystalline magnetic
anisotropy, only shape and strain-induced anisotropies need to be
considered \cite{Bihler:PRB:78:2008}. The angular dependence of the
total free energy density $F$ within the plane of such a uniaxially
strained ferromagnetic film shows a $180^{\circ}$ periodicity, where
maxima of $F$ correspond to magnetically hard directions, and accordingly
minima to magnetically easy directions. In the Stoner-Wohlfarth (SW)
model \cite{Stoner:PhilosTransRSocLondon:240:1948}, the magnetization
orientation can be calculated from $F$, since in equilibrium $\mathbf{M}$
aligns along a local minimum of $F$. Regarding the in-plane magnetoelastic
contribution to $F$ \cite{Chikazumi:book:1997}
\begin{equation}
F_{\textrm{me}}=\frac{3}{2}\bar{\lambda}\left(c_{12}-c_{11}\right)\left[\epsilon_{x}\left(\alpha_{x}^{2}-\frac{1}{3}\right)+\epsilon_{y}\left(\alpha_{y}^{2}-\frac{1}{3}\right)\right],
\end{equation}
with the elastic stiffness constants $c_{ij}$ \cite{Lee:RPP:18:1955}
and the direction cosines of the magnetization $\alpha_{x}$ and $\alpha_{y}$
with respect to $\mathbf{x}$ and $\mathbf{y}$, respectively, it
is important to note that for nickel $\frac{3}{2}\bar{\lambda}\left(c_{12}-c_{11}\right)>0$.
Therefore, in the absence of external magnetic fields, the magnetic
easy axis is oriented parallel to compressive strain ($\epsilon<0$)
and orthogonal to tensile strain ($\epsilon>0$). Consequently, an
applied voltage $V_{\mathrm{p}}>\unit[0]{V}$ results in a magnetic
easy axis and thus an equilibrium magnetization orientation along
$\mathbf{x}$ {[}cf. red contour in Fig.~\ref{fig:concept}(b){]},
while accordingly for $V_{\mathrm{p}}<\unit[0]{V}$ the magnetization
is oriented along $\mathbf{y}$ {[}cf. blue contour in Fig.~\ref{fig:concept}(b){]}.
Hence, we expect a $90^{\circ}$ rotation of the easy axis and thus
the magnetization orientation upon inverting the polarity of $V_{\mathrm{p}}$.

Figure~\ref{fig:concept}(c) shows calculated magnetization curves,
normalized to the saturation magnetization $M_{\mathrm{s}}$, i.e.,
$M=\left(\mathbf{M}\cdot\mathbf{x}\right)/M_{\mathrm{s}}$, as a function
of the external magnetic field magnitude $H$ at fixed voltages
$V_{\mathrm{p}}$. The curves are calculated using a single-domain
SW model for the external magnetic field $\mathbf{H}$ applied along
$\mathbf{x}$ {[}cf.~Fig.~\ref{fig:concept}(a){]}, as this is the
case for all experimental data presented below. The SW model relies
on the coherent rotation of a single homogeneous magnetic domain (macrospin).
For $V_{\mathrm{p}}>\unit[0]{V}$, the $\mathbf{x}$ direction coincides
with the easy axis {[}e.a., see Fig.~\ref{fig:concept}(b){]} and
thus the corresponding $M\left(H\right)$ loop {[}red curve in Fig.~\ref{fig:concept}(c){]}
exhibits a rectangular, hysteretic shape \cite{Morrish:book:2001},
which indicates an abrupt $180^{\circ}$ magnetization switching (i.e.,
a discontinuous magnetization reversal) at the coercive field $H_{\mathrm{c}}$.
Contrarily, the $\mathbf{x}$ direction is magnetically hard for $V_{\mathrm{p}}<\unit[0]{V}$,
and hence the blue magnetization curve in Fig.~\ref{fig:concept}(c)
exhibits no magnetic hysteresis as typically observed for hard-axis
loops, which indicates a magnetization-reversal process via continuous
rotation.

To determine the static magnetic properties of the Ni thin film/actuator
hybrid, we employ spatially resolved magneto-optical Kerr effect (MOKE)
imaging. More precisely, we perform longitudinal MOKE spectroscopy,
which detects the projection $M=\mathbf{M}\cdot\mathbf{x}$ of the
magnetization onto the magnetic field direction $\mathbf{H}\parallel\mathbf{x}$.
All data were recorded at room temperature. Our MOKE setup is equipped
with a high power red light emitting diode (center wavelength $\lambda=\unit[627]{nm}$).
A slit aperture and a Glan Thompson polarizing prism yield an illumination
path with $s$-polarized incident light. After reflecting off the
sample, the light passes through a quarter wave plate to remove the
ellipticity, and then transmits through a second Glan Thompson polarizing
prism close to extinction serving as the analyzer. The Kerr signal
is then focused by an objective lens and recorded with a CCD camera
with a pixel size of $\unit[10]{\mu m}\times\unit[10]{\mu m}$. While
the setup has a rather low spatial resolution of several micrometers,
it allows to image samples with lateral dimensions of several $\unit{mm^{2}}$
in a single-shot experiment. Such a large field of view is mandatory
to investigate the piezo-induced $\mathbf{M}\left(V_{\mathrm{p}}\right)$,
since the actuator electrodes are about $\unit[10]{\mu m}$ wide,
and the active piezoelectric regions are about $\unit[100]{\mu m}$
wide.

To enable AMR measurements simultaneously to MOKE, we contacted the
Ni film on top in four-point geometry {[}see Fig.~\ref{fig:concept}(a){]}.
All AMR data shown in the following were recorded with a constant
bias current $\mathbf{I}$ parallel to $\mathbf{H}\parallel\mathbf{x}$. The longitudinal
resistance in a single-domain model is given by \cite{McGuire:IEEETransMagn:11:1975}
\begin{equation}
R=R_{\perp}+\left(R_{\parallel}-R_{\perp}\right)\cos^{2}\beta,\label{eq:AMR}
\end{equation}
where $\beta$ is the angle between $\mathbf{M}$ and $\mathbf{I}$,
and $R_{\perp}$ and $R_{\parallel}$ are the resistances at $\mathbf{I}\perp\mathbf{M}$
and $\mathbf{I}\parallel\mathbf{M}$, respectively.

As discussed in more detail in the results section, we apply several image-processing
procedures to extract the relevant magnetic information. On the one
hand, we use the difference-image technique, i.e., the digital subtraction
of two images, to enhance the magneto-optical contrast and to exclude
any non-magnetic signal contributions. To this end, a reference image
is recorded in a magnetically saturated state and subtracted from
subsequent images \cite{Schmidt:IEEETransMagn:21:1985}. On the other
hand, for a quantitative magnetization analysis we normalize the observed
image contrast. To this end, we define a region of interest (ROI),
which corresponds to the region covered with Ni. Within this
ROI, we integrate over all pixels of the CCD-camera image and normalize
the resulting value with respect to the ones related to the two opposite
single-domain saturation states. This evaluation yields an effective
averaged magnetization $-1\leq M/M_{\mathrm{s}}\leq1$. The latter
can be considered as an effective macrospin, in which any microscopic
magnetic texture has been averaged out. It will be referred to as
{}``integrated MOKE loop'' in the following.

\section{Results and discussion}

\begin{figure}
\begin{centering}
\includegraphics[width=1\columnwidth]{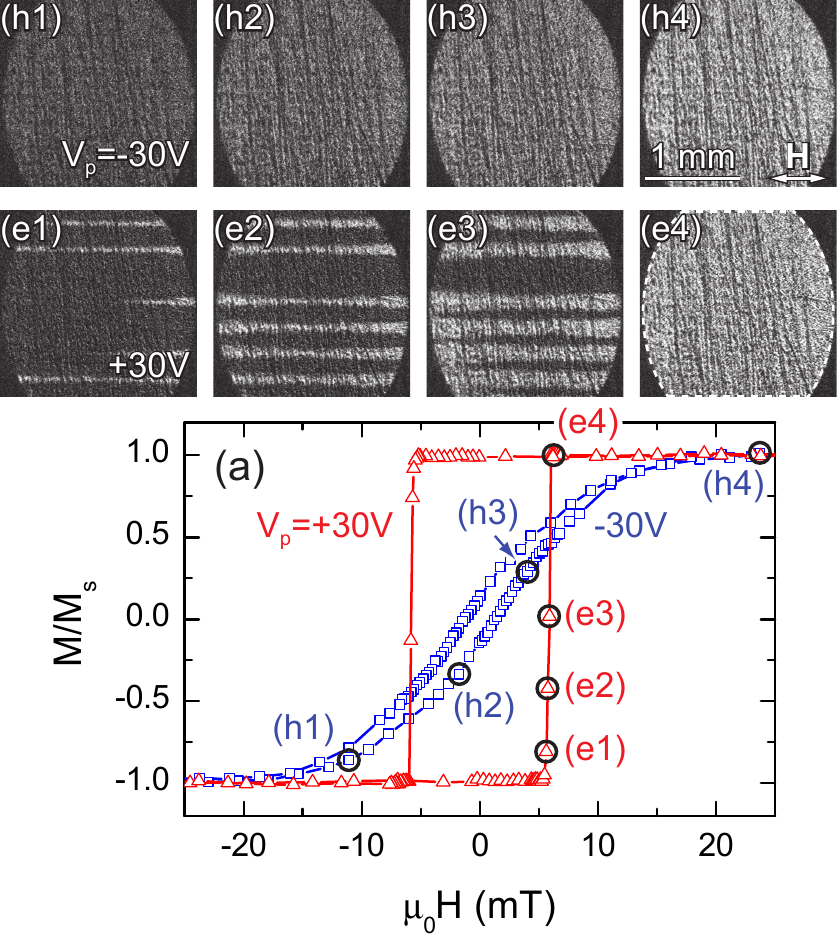}
\par\end{centering}

\caption{Magnetic domain evolution as a function of the external magnetic field
magnitude. For $V_{\mathrm{p}}=\unit[-30]{V}$, the magnetic hard
axis is along $\mathbf{H}\parallel\mathbf{x}$, while for $V_{\mathrm{p}}=\unit[+30]{V}$,
the easy axis is along $\mathbf{H}\parallel\mathbf{x}$. (a) Integrated
MOKE $M\left(H\right)$ loops for $V_{\mathrm{p}}=\unit[-30]{V}$
(open blue squares) and $V_{\mathrm{p}}=\unit[+30]{V}$ (open red
triangles), using the region of interest depicted by the dashed white
contour in (e4). The lines are guides to the eye. The open black circles
mark the magnetic fields at which the MOKE images (h1)--(h4) and (e1)--(e4)
were taken. Hereby, (h1)--(h4) were recorded for $V_{\mathrm{p}}=\unit[-30]{V}$
($\mathbf{H}$ parallel to the magnetic hard axis) and (e1)--(e4)
were recorded for $V_{\mathrm{p}}=\unit[+30]{V}$ ($\mathbf{H}$ parallel
to the easy axis).\label{fig:MH_DomainEvol}}
\end{figure}
In a first series of experiments, we studied the magnetic domain evolution
at constant strain, i.e., we recorded MOKE images at constant voltages
$V_{\mathrm{p}}$ as a function of the external magnetic field magnitude
$H$ for fixed magnetic field orientation $\mathbf{H}\parallel\mathbf{x}$.
We refer to these experiments as $M\left(H\right)$ measurements.
We applied a fixed voltage $V_{\mathrm{p}}$ to the actuator, swept
the magnetic field to $\mu_{0}H=\unit[-50]{mT}$ to prepare the magnetization
in a single-domain, negative saturation state, and acquired a reference
image. Subsequently, the magnetic field was increased in steps, and
a MOKE image was recorded at every field value. The corresponding
domain evolution (obtained after subtraction of the reference image)
and the integrated MOKE loops for $V_{\mathrm{p}}=\unit[-30]{V}$
and $V_{\mathrm{p}}=\unit[+30]{V}$ are shown in Fig.~\ref{fig:MH_DomainEvol}.
The integrated MOKE loops in Fig.~\ref{fig:MH_DomainEvol}(a) clearly
resemble the SW simulations in Fig.~\ref{fig:concept}(c), such that
at $V_{\mathrm{p}}=\unit[-30]{V}$ we obtain a smooth and continuous
$M\left(H\right)$ loop for a magnetic hard axis along $\mathbf{H}\parallel\mathbf{x}$
(open blue squares), while $V_{\mathrm{p}}=\unit[+30]{V}$ yields
a magnetic easy axis along $\mathbf{x}$ and thus results in a rectangular-shaped
$M\left(H\right)$ loop with discontinuous magnetization-reversal
processes (open red triangles).

We start the discussion with the data recorded at $V_{\mathrm{p}}=\unit[-30]{V}$,
i.e., for a magnetic hard axis along $\mathbf{x}$. Figures~\ref{fig:MH_DomainEvol}(h1)--(h4)
show MOKE images for an upsweep of the external magnetic field obtained
at magnetic field values $\mathrm{h1}=\unit[-10.5]{mT}$, $\mathrm{h2}=\unit[-1.5]{mT}$,
$\mathrm{h3}=\unit[4.0]{mT}$, and $\mathrm{h4}=\unit[23.0]{mT}$,
marked by circles in the corresponding $M\left(H\right)$ loop presented
graphically by square symbols in Fig.~\ref{fig:MH_DomainEvol}(a).
In Figs.~\ref{fig:MH_DomainEvol}(h1)--(h4), the spatially resolved
MOKE intensity is homogeneous over the whole Ni film, continuously
changing from black to white with increasing magnetic field strength
from negative to positive saturation. Hence, the sample is uniformly
magnetized throughout the magnetization-reversal process, suggesting
coherent and continuous magnetization rotation. Clearly, a single-domain
SW approach is appropriate to model this behavior. In contrast, the
magnetization reversal along the magnetic easy axis at $V_{\mathrm{p}}=\unit[+30]{V}$
is depicted in Figs.~\ref{fig:MH_DomainEvol}(e1)--(e4) at magnetic
fields close to the coercive field {[}cf. $M\left(H\right)$ loop
shown by open red triangles in Fig.~\ref{fig:MH_DomainEvol}(a){]}.
As apparent, magnetic domains nucleate and gradually propagate until
the magnetization-reversal process is finally complete in Fig.~\ref{fig:MH_DomainEvol}(e4).
For such a domain-driven magnetization-reversal process the simple
SW single macrospin approach appears inadequate, since the magnetization-reversal
process in $M\left(H\right)$ measurements is usually modeled by combining
coherent rotation and domain-wall nucleation and/or unpinning \cite{Florczak:PRB:44:1991}.
More precisely, in ferromagnetic thin films with uniaxial anisotropy,
a magnetic-field induced magnetization reversal is determined by coherent
rotation when the external magnetic field is oriented close to the
magnetic hard axis, which is thus in good agreement with the SW model.
In contrast, when the magnetic field is oriented along the magnetic
easy axis, the abrupt magnetization reversal is caused by domain-wall
effects, and thus proceeds via noncoherent switching. For other orientations,
the magnetization first continuously reorients by coherent rotation,
and then switches discontinuously and noncoherently (for details,
see, e.g., Ref.~\onlinecite{Yan:PRB:63:2001}). In other words, in view
of the domain pattern in Figs.~\ref{fig:MH_DomainEvol}(e1)--(e4),
the macrospin model used in the literature for the magnetization reorientation
$M\left(E\right)$ as a function of electric field at fixed external
magnetic field magnitude \cite{Weiler:NJP:11:2009} only represents
a first-order approximation.

\begin{figure}
\begin{centering}
\includegraphics[width=1\columnwidth]{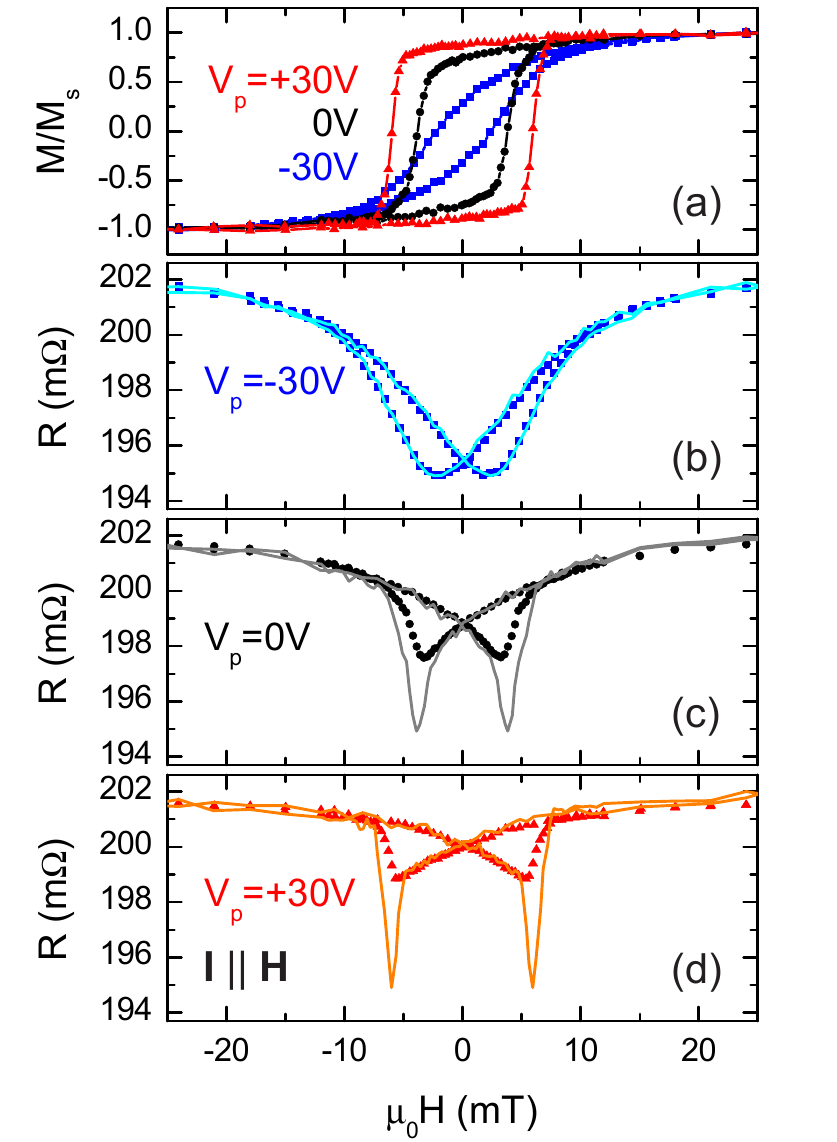}
\par\end{centering}

\caption{(a) Integrated MOKE $M\left(H\right)$ loops for $V_{\mathrm{p}}=\unit[-30]{V}$
(blue squares), $V_{\mathrm{p}}=\unit[0]{V}$ (black circles), and
$V_{\mathrm{p}}=\unit[+30]{V}$ (red triangles). Simultaneously measured
(full symbols) and simulated (solid lines) AMR $R\left(H\right)$
loops for $V_{\mathrm{p}}=\unit[-30]{V}$ (b), $V_{\mathrm{p}}=\unit[0]{V}$
(c), and $V_{\mathrm{p}}=\unit[+30]{V}$ (d). Apart from deviations
in the vicinity of the coercive fields for $V_{\mathrm{p}}=\unit[0]{V}$
and $V_{\mathrm{p}}=\unit[+30]{V}$, the simulations based on a pseudo-macrospin
model are in very good agreement with the measurements.\label{fig:MH_RH}}
\end{figure}
To examine the validity and limitations of the macrospin model
for $M\left(E\right)$ in more detail, we now address the AMR recorded simultaneously
to the MOKE data, referred to as $R\left(H\right)$ measurements.
Figure~\ref{fig:MH_RH}(a) again depicts integrated MOKE loops along
a magnetic hard axis ($V_{\mathrm{p}}=\unit[-30]{V}$, full blue squares),
with zero applied stress ($V_{\mathrm{p}}=\unit[0]{V}$, full black
circles), and along a magnetic easy axis ($V_{\mathrm{p}}=\unit[+30]{V}$,
full red triangles). The corresponding AMR loops, represented by full
symbols, are shown in Figs.~\ref{fig:MH_RH}(b), (c), and (d) for
$V_{\mathrm{p}}=\unit[-30]{V}$, $V_{\mathrm{p}}=\unit[0]{V}$, and
$V_{\mathrm{p}}=\unit[+30]{V}$, respectively. To quantitatively simulate
the evolution of $R\left(H\right)$ in a macrospin-type
SW model, we determined an effective, average magnetization orientation
$\beta$ from the $M\left(H\right)$ loops in Fig.~\ref{fig:MH_RH}(a)
via $\cos\beta=M/M_{\mathrm{s}}$ (cf.~Ref.~\onlinecite{Brandlmaier:JAP:2011}). To this end, we use the effective,
averaged magnetization orientation in the ROI as a pseudo-macrospin.
It should be emphasized at this point that this pseudo-macrospin corresponds
to a magnetization orientation $\beta$ averaged over differently
oriented magnetic domains, see Figs.~\ref{fig:MH_DomainEvol}(e1)--(e4).
Equation~\eqref{eq:AMR} with $R_{\parallel}=\unit[201.9]{m\Omega}$
and $R_{\perp}=\unit[194.9]{m\Omega}$ then yields the solid lines
in Figs.~\ref{fig:MH_RH}(b), (c), and (d). As evident from the figure,
the AMR calculated using the macrospin model accurately reproduces
the measured AMR for $\mathbf{H}$ parallel to a hard axis ($V_{\mathrm{p}}=\unit[-30]{V}$).
For $V_{\mathrm{p}}=\unit[0]{V}$ and $V_{\mathrm{p}}=\unit[+30]{V}$,
$\mathbf{H}$ is along an axis with increasingly easy character and
we observe an increasing deviation of the AMR simulation from the
AMR experiment---however only for $H$ in the vicinity of the coercive
fields {[}see Figs.~\ref{fig:MH_RH}(c) and (d){]}. Hence, the AMR experiments
corroborate the notion that the pseudo-macrospin is not adequate in
the case of substantial microscopic domain formation, i.e., close
to the coercive fields. Interestingly, however, the simulations yield
very good agreement with the experimental results for all other magnetic
field values. In summary, the magnetization reversal at constant strain
in our multifunctional hybrid systems can be modeled in very good
approximation using a pseudo-macrospin type of approach, except for
a small range around $H\approx H_{\mathrm{c}}$ with substantial domain
formation.

\begin{figure}
\begin{centering}
\includegraphics[width=1\columnwidth]{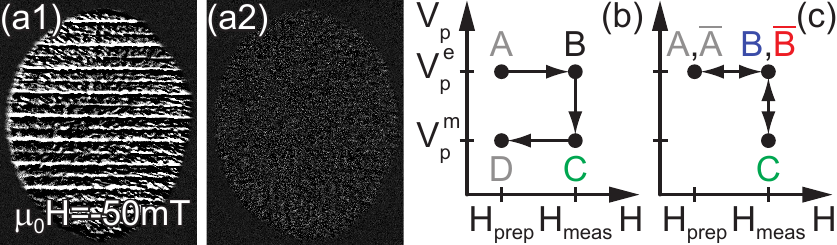}
\par\end{centering}

\caption{(a) Elastic strain-induced Kerr contrast patterns, recorded with the
Ni film magnetized to saturation via the application of $\mu_{0}H=\unit[-50]{mT}$
at $V_{\mathrm{p}}=\unit[+30]{V}$. (a1) Difference image after a
sweep to $V_{\mathrm{p}}=\unit[-30]{V}$ and (a2) back to $V_{\mathrm{p}}=\unit[+30]{V}$.
(b), (c) Schematic illustrations of measurement sequences used to
record magnetic Kerr images as a function of $V_{\mathrm{p}}$ without
the strain-induced signal contributions.\label{fig:biref}}
\end{figure}
In a second set of experiments we address the voltage control of the
magnetization orientation. To this end, we record MOKE images and
the AMR as a function of the voltage $V_{\mathrm{p}}$ at constant
external magnetic bias field $\mathbf{H}$. However, it turns out
that the different strain states at different $V_{\mathrm{p}}$ also
give raise to a Kerr signal, as illustrated in Fig.~\ref{fig:biref}.
In these experiments, we aligned a magnetic easy axis along $\mathbf{H}\parallel\mathbf{x}$
by applying $V_{\mathrm{p}}=\unit[+30]{V}$, and initially magnetized
the sample to saturation in a single-domain state by
sweeping the magnetic field to $\mu_{0}H=\unit[-50]{mT}$. After recording
a reference image, we swept the voltage to $V_{\mathrm{p}}=\unit[-30]{V}$,
keeping $\mu_{0}H=\unit[-50]{mT}$ constant. Figure~\ref{fig:biref}(a1)
shows the difference image at $V_{\mathrm{p}}=\unit[-30]{V}$ with
respect to the reference at $V_{\mathrm{p}}=\unit[+30]{V}$. After
sweeping the voltage back to $V_{\mathrm{p}}=\unit[+30]{V}$, the
Kerr difference contrast pattern completely vanishes {[}Fig.~\ref{fig:biref}(a2){]},
indicating the reversibility of this process. Evidently, the observed
Kerr contrast cannot be of magnetic origin, since the magnetic field
$\mu_{0}H=\unit[-50]{mT}$ is way large enough to ensure magnetic
saturation at any $V_{\mathrm{p}}$, such that neither the magnetization
orientation nor the magnitude are subject to magnetoelastic modifications.
The Kerr contrast thus must be of non-magnetic origin. Two mechanisms
may account for these strain-induced contrast changes. First, the
Poisson ratios of the PZT piezoelectric layers and the interdigitated
electrodes differ. Thus, the strain induced by $V_{\mathrm{p}}\neq\unit[0]{V}$
will be slightly inhomogeneous in the $\mathbf{x}-\mathbf{y}$ plane
of the actuator, leading to local surface corrugations above the electrodes
and thus to a modified intensity of the reflected light. Second, the
PMMA layer in between piezo and Ni exhibits strain-induced birefringence,
e.g., photoelastic birefringence \cite{Ohkita:ApplPhysA:81:2005},
which also results in a Kerr contrast.

To nevertheless extract the magnetic Kerr signal contributions, we
apply two more elaborate measurement sequences. The basic sequence
is schematically illustrated in Fig.~\ref{fig:biref}(b). We magnetize
the sample to a single-domain state by applying a magnetic preparation
field $\mu_{0}H_{\mathrm{prep}}=\unit[-50]{mT}$ well exceeding the
saturation field along the easy axis at $V_{\mathrm{p}}^{\mathrm{e}}=\unit[+30]{V}$
(point $\mathrm{A}$), then sweep the magnetic field to the measurement
bias field $H_{\mathrm{meas}}$ (point $\mathrm{B}$), in turn sweep
the voltage to the measurement voltage $V_{\mathrm{p}}^{\mathrm{m}}$
(point $\mathrm{C}$), and finally go back to the preparation field
$H_{\mathrm{prep}}$ (point $\mathrm{D}$). A Kerr image is
acquired at each point. Subsequently, we subtract the images corresponding
to equal strain states, such that the resulting images $\mathrm{B}-\mathrm{A}$
and $\mathrm{C}-\mathrm{D}$ exhibit only contrast of magnetic origin.
Hence, this procedure allows to image the evolution of $\mathbf{M}$
as a function of strain. To also investigate the reversibility of
the voltage-induced magnetization changes, we modify the sequence
as sketched in Fig.~\ref{fig:biref}(c). After sweeping the voltage
to the measurement voltage $V_{\mathrm{p}}^{\mathrm{m}}$ (point $\mathrm{C}$),
it is cycled back to $V_{\mathrm{p}}^{\mathrm{e}}=\unit[+30]{V}$
(point $\overline{\mathrm{B}}$), and finally the magnetic field is
returned to $H_{\mathrm{prep}}$ (point $\overline{\mathrm{A}}$).
The difference images $\mathrm{B}-\mathrm{A}$ and $\overline{\mathrm{B}}-\overline{\mathrm{A}}$
then reveal the degree of reversibility.

\begin{figure}
\begin{centering}
\includegraphics[width=1\columnwidth]{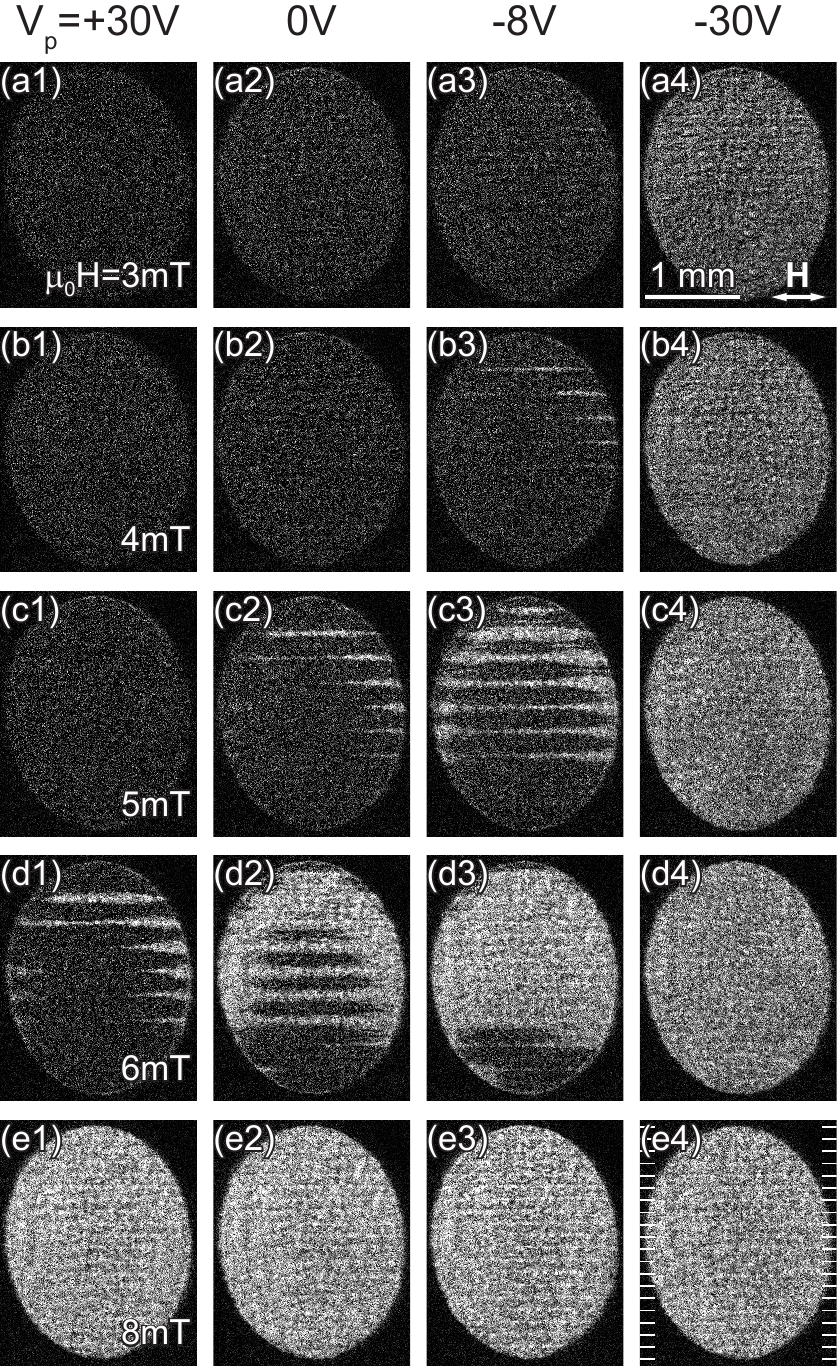}
\par\end{centering}

\caption{Evolution of the magnetization reorientation as a function of the
voltage $V_{\mathrm{p}}$ at different, fixed magnetic bias fields
after preparing a magnetic single-domain state at $\mu_{0}H_{\mathrm{prep}}=\unit[-50]{mT}$.
The magnetic Kerr images are obtained using the sequence depicted
in Fig.~\ref{fig:biref}(b) (images in the first column show difference
images $\mathrm{B}-\mathrm{A}$, all other images show difference
images $\mathrm{C}-\mathrm{D}$). In the vicinity of the coercive
field {[}$\mu_{0}H_{\mathrm{c}}\approx\unit[6]{mT}$ at $V_{\mathrm{p}}=\unit[+30]{V}$,
cf.~Fig.~\ref{fig:MH_RH}(a){]} for $\mu_{0}H_{\mathrm{meas}}=\unit[4]{mT}$
{[}(b1)--(b4){]}, $\unit[5]{mT}$ {[}(c1)--(c4){]}, and $\unit[6]{mT}$
{[}(d1)--(d4){]}, the voltage-controlled magnetization-reorientation
process evidently occurs via magnetic domain nucleation and propagation,
and finally attains a single-domain state at $V_{\mathrm{p}}=\unit[-30]{V}$.
For other magnetic field strengths, e.g., $\mu_{0}H_{\mathrm{meas}}=\unit[3]{mT}$
{[}(a1)--(a4){]} and $\mu_{0}H_{\mathrm{meas}}=\unit[8]{mT}$ {[}(e1)--(e4){]},
the Kerr image contrast changes homogeneously as a function of $V_{\mathrm{p}}$,
i.e., the magnetization rotates coherently during the $V_{\mathrm{p}}$
sweep. The white lines in (e4) indicate the position of the actuator
electrodes.\label{fig:ME-Img}}
\end{figure}
Figure~\ref{fig:ME-Img} shows the magnetic Kerr images obtained
using the sequence depicted in Fig.~\ref{fig:biref}(b) as a function
of the voltage $V_{\mathrm{p}}$ at different fixed magnetic bias
fields $\mu_{0}H_{\mathrm{meas}}=\unit[3]{mT}$, $\unit[4]{mT}$,
$\unit[5]{mT}$, $\unit[6]{mT}$, and $\unit[8]{mT}$, depicted in
Figs.~\ref{fig:ME-Img}(a1)--(a4), (b1)--(b4), (c1)--(c4), (d1)--(d4),
and (e1)--(e4), respectively. Hereby, the images (a)--(c) are recorded
slightly below the coercive field, while (d) ($\mu_{0}H_{\mathrm{meas}}=\unit[6]{mT}$)
is directly at the coercive field {[}$\mu_{0}H_{\mathrm{c}}\approx\unit[6]{mT}$
at $V_{\mathrm{p}}=\unit[+30]{V}$, cf.~Fig.~\ref{fig:MH_RH}(a){]}.
For the latter {[}Fig.~\ref{fig:ME-Img}(d1){]}, domain nucleation
already starts without changing $V_{\mathrm{p}}$. The difference
images $\mathrm{B}-\mathrm{A}$ shown in the first column of Fig.~\ref{fig:ME-Img}
are acquired at the preparation voltage $V_{\mathrm{p}}^{\mathrm{e}}=\unit[+30]{V}$
after a magnetic field sweep from $\mu_{0}H_{\mathrm{prep}}=\unit[-50]{mT}$
to the bias field $H_{\mathrm{meas}}$. The difference images $\mathrm{C}-\mathrm{D}$
in the latter three columns result from a consecutive application
of the measurement sequence with different measurement voltages $V_{\mathrm{p}}^{\mathrm{m}}=\unit[0]{V}$,
$\unit[-8]{V}$, and $\unit[-30]{V}$. As evident from Figs.~\ref{fig:ME-Img}(a1),
(b1), and (c1), no magnetic contrast is yet visible at $H_{\mathrm{meas}}$,
indicating a uniform, single-domain magnetization along the initial
magnetic field orientation $\mu_{0}H_{\mathrm{prep}}<\unit[0]{mT}$,
i.e., antiparallel to the bias magnetic field orientation $\mu_{0}H_{\mathrm{meas}}>\unit[0]{mT}$.
Upon gradually decreasing $V_{\mathrm{p}}$, a noncoherent magnetization-reorientation
process sets in for magnetic fields close to the coercive field {[}see
Figs.~\ref{fig:ME-Img}(b3) and (c2){]} via magnetic domain nucleation
and propagation, until the process is complete at $V_{\mathrm{p}}=\unit[-30]{V}$,
as shown in the images in the last column. We note that the domain
nucleation preferably proceeds on top of the electrodes, which we
attribute to the slight strain inhomogeneities discussed in the context
of Fig.~\ref{fig:biref}(a1). The final image contrast after the
magnetization reorientation is homogeneously white {[}Figs.~\ref{fig:ME-Img}(a4)
to (e4){]}, evidencing a magnetic single-domain state.

Figures~\ref{fig:ME-Img}(a1)--(a4), (b1)--(b4), (c1)--(c4), and
(e1)--(e4) show a fully voltage-controlled magnetization reorientation
from an initial magnetic single-domain state to a final single-domain
state. As apparent from the Kerr images in Fig.~\ref{fig:ME-Img},
for externally applied magnetic fields close to $H_{\mathrm{c}}$
the magnetization-reorientation process evolves via domain nucleation
and propagation. In contrast, for other magnetic field strengths {[}$\mu_{0}H_{\mathrm{meas}}<\unit[4]{mT}$
and $\mu_{0}H_{\mathrm{meas}}>\unit[7]{mT}$, see Figures~\ref{fig:ME-Img}(a1)--(a4)
and (e1)--(e4){]}) the Kerr image contrast changes homogeneously as
a function of $V_{\mathrm{p}}$, i.e., the magnetization rotates coherently
during the $V_{\mathrm{p}}$ sweep.

\begin{figure}
\begin{centering}
\includegraphics[width=1\columnwidth]{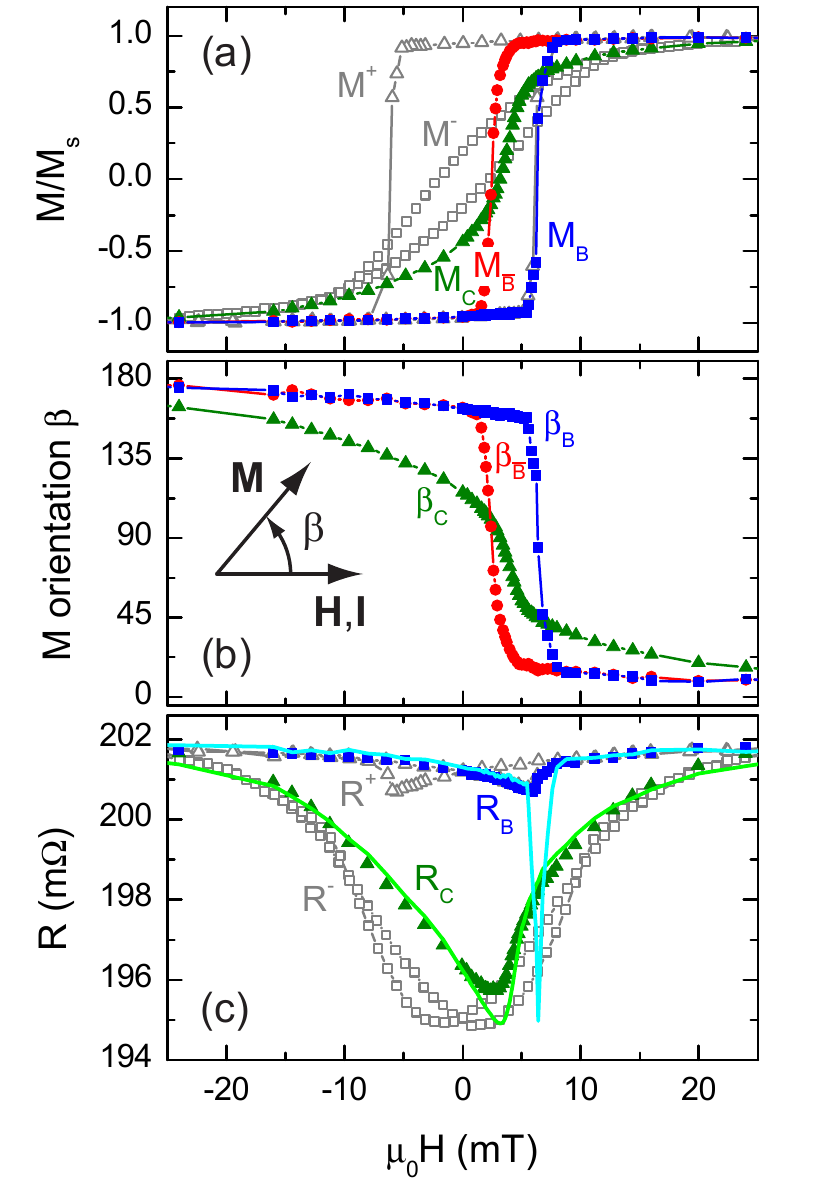}
\par\end{centering}

\caption{Voltage-dependent $M$ and $R$ measurements. The panels show data
acquired at the points $\mathrm{B}$ (full blue squares), $\overline{\mathrm{B}}$
(full red circles), and $\mathrm{C}$ (full green triangles) as a
function of the measurement field $H=H_{\mathrm{meas}}$ in the sequences
depicted in Figs.~\ref{fig:biref}(b) and (c). (a) Integrated MOKE
$M\left(V_{\mathrm{p}}\right)$ curves obtained from the respective
difference images. For comparison, the open gray squares and open
gray triangles depict the $M\left(H\right)$ loops for $V_{\mathrm{p}}=\unit[-30]{V}$
($\mathrm{M}^{-}$) and $\unit[+30]{V}$ ($\mathrm{M}^{+}$), respectively.
(b) Macrospin magnetization orientation $\beta$ with the angle $\beta$
between $\mathbf{M}$ and $\mathbf{I}$, calculated from the measured
data displayed in (a) using the pseudo-macrospin model. (c) Corresponding
AMR $R\left(V_{\mathrm{p}}\right)$ curves (symbols) and simulations
(lines) using the simultaneously recorded $M\left(V_{\mathrm{p}}\right)$
curves {[}(a) and (b){]} and the AMR parameters given in the text showing
very good overall agreement with the experiment. Hereby, the blue
and green solid lines illustrate the simulated evolution of $R_{\mathrm{B}}\left(H\right)$
and $R_{\mathrm{C}}\left(H\right)$, respectively. In analogy to (a),
the $R\left(H\right)$ loops for $V_{\mathrm{p}}=\unit[-30]{V}$ ($\mathrm{R}^{-}$)
and $\unit[+30]{V}$ ($\mathrm{R}^{+}$) are additionally depicted
by open gray squares and open gray triangles, respectively.\label{fig:ME-RE}}
\end{figure}
To quantitatively evaluate the Kerr images recorded as a function
of $V_{\mathrm{p}}$, we finally test the applicability of a macrospin
picture for the description of $M\left(V_{\mathrm{p}}\right)$, and
address the reversibility of the voltage-induced magnetization reorientation.
To this end, we consecutively applied the two measurement sequences
illustrated in Figs.~\ref{fig:biref}(b) and (c) with different measurement
bias fields $H_{\mathrm{meas}}$. More precisely, for each $H_{\mathrm{meas}}$
we first applied the basic sequence shown in Fig.~\ref{fig:biref}(b)
with $\mu_{0}H_{\mathrm{prep}}=\unit[-50]{mT}$, $V_{\mathrm{p}}^{\mathrm{e}}=\unit[+30]{V}$,
and $V_{\mathrm{p}}^{\mathrm{m}}=\unit[-30]{V}$. Then we applied
the modified sequence shown in Fig.~\ref{fig:biref}(c) at the same
$H_{\mathrm{meas}}$. At each point, a Kerr image is acquired and
the resistance is recorded. We refer to these experiments as $M\left(V_{\mathrm{p}}\right)$
and $R\left(V_{\mathrm{p}}\right)$ measurements, respectively. The
evolution of the corresponding integrated MOKE $M\left(V_{\mathrm{p}}\right)$
signals obtained from the difference images $\mathrm{B}-\mathrm{A}$,
$\overline{\mathrm{B}}-\overline{\mathrm{A}}$, and $\mathrm{C}-\mathrm{D}$
as a function of the measurement field $H=H_{\mathrm{meas}}$ is shown
in Fig.~\ref{fig:ME-RE}(a) and referred to as $M_{\mathrm{B}}\left(H\right)$
(full blue squares), $M_{\overline{\mathrm{B}}}\left(H\right)$ (full
red circles), and $M_{\mathrm{C}}\left(H\right)$ (full green triangles),
respectively. We again determined the magnetization orientation $\beta$
in a macrospin approximation, as displayed in Fig.~\ref{fig:ME-RE}(b).
The corresponding AMR $R\left(V_{\mathrm{p}}\right)$ curves simultaneously
measured with the Kerr images at the points $\mathrm{B}$ and $\mathrm{C}$
are depicted in Fig.~\ref{fig:ME-RE}(c) by full blue squares and
full green triangles and referred to as $R_{\mathrm{B}}\left(H\right)$
and $R_{\mathrm{C}}\left(H\right)$, respectively. For comparison,
we also included the integrated MOKE $M\left(H\right)$ and AMR $R\left(H\right)$
data recorded as a function of the external magnetic field at constant
$V_{\mathrm{p}}=\unit[-30]{V}$ and $\unit[+30]{V}$ as open gray
squares {[}denoted as $\mathrm{M}^{-}$ and $\mathrm{R}^{-}$ in Figs.~\ref{fig:ME-RE}(a)
and (c), respectively{]} and open gray triangles ($\mathrm{M}^{+}$
and $\mathrm{R}^{+}$), respectively.

Using the $M_{\mathrm{B}}\left(H\right)$ and $M_{\mathrm{C}}\left(H\right)$
curves {[}Fig.~\ref{fig:ME-RE}(a){]} recorded simultaneously with
$R_{\mathrm{B}}\left(H\right)$ and $R_{\mathrm{C}}\left(H\right)$,
respectively, we again simulate the AMR with Eq.~\eqref{eq:AMR}
and the values of the parameters $R_{\parallel}$ and $R_{\perp}$
given above. The AMR curves thus calculated for $R_{\mathrm{B}}\left(H\right)$
and $R_{\mathrm{C}}\left(H\right)$ are displayed by the solid blue
and green lines in Fig.~\ref{fig:ME-RE}(c), respectively. Evidently,
the measurement and simulation for both $R_{\mathrm{B}}\left(H\right)$
and $R_{\mathrm{C}}\left(H\right)$ are in very good agreement, with
the exception of a narrow region around $H_{\mathrm{c}}$, where the
macrospin model fails to adequately describe the experiment, in full
consistency with the above findings for $M\left(H\right)$ and $R\left(H\right)$
measurements (cf.~Fig.~\ref{fig:MH_RH}). Therefore, except for
a small magnetic-field range close to $H_{\mathrm{c}}$ we can describe
the voltage-induced magnetization changes in very good approximation
in a simple macrospin model.

As we now have demonstrated the validity of the macrospin approach
also for the description of $M\left(V_{\mathrm{p}}\right)$ and $R\left(V_{\mathrm{p}}\right)$ measurements,
we can consistently model the evolution of the magnetization orientation
as a function of the voltage $V_{\mathrm{p}}$ and the external magnetic
field $H$ (Fig.~\ref{fig:ME-RE}). We start the discussion
with the evolution of $M_{\mathrm{B}}\left(H\right)$ {[}full blue
squares in Fig.~\ref{fig:ME-RE}(a){]}, which coincides with the
$M\left(H\right)$ loop recorded at $V_{\mathrm{p}}=\unit[+30]{V}$.
The corresponding macrospin magnetization orientation $\beta_{\mathrm{B}}$ is
initially aligned along $180^{\circ}$ for large negative external
magnetic field {[}full blue squares in Fig.~\ref{fig:ME-RE}(b){]},
continuously rotates to $\approx160^{\circ}$ with increasing magnetic
field strength, then abruptly switches into a direction close to $\beta_{\mathrm{B}}=0^{\circ}$
at the magnetic coercive field, and then continuously rotates towards
$0^{\circ}$, the orientation of the external magnetic field. For
external magnetic measurement fields $\mu_{0}H_{\mathrm{meas}}\lesssim\unit[0]{mT}$,
the influence of the Zeeman contribution to the total free energy
density in the film plane $F=F_{\mathrm{Zeeman}}+F_{\textrm{me}}$
decreases with decreasing absolute value of the external magnetic
field, which results in an increasingly dominating magnetoelastic
anisotropy contribution. Hence, the magnetization orientation cannot
be modified at $\mu_{0}H_{\mathrm{prep}}=\unit[-50]{mT}$ by application
of $V_{\mathrm{p}}$ and can be increasingly rotated to about $\Delta\beta=50^{\circ}$
at $\mu_{0}H=\unit[0]{mT}$ by changing $V_{\mathrm{p}}=\unit[+30]{V}\rightarrow\unit[-30]{V}$
{[}$\beta_{\mathrm{B}}\left(\mu_{0}H=\unit[0]{mT}\right)\approx165^{\circ}$
and $\beta_{\mathrm{C}}\left(\mu_{0}H=\unit[0]{mT}\right)\approx115^{\circ}${]}.
In this magnetic field range, $M_{\mathrm{B}}\left(H\right)$ and
$M_{\overline{\mathrm{B}}}\left(H\right)$ fully coincide ($\mathbf{M}_{\mathrm{B}}\parallel\mathbf{M}_{\overline{\mathrm{B}}}$),
i.e., the voltage-induced magnetization reorientation is fully reversible.
We would like to emphasize that this implies a continuous and reversible
magnetization rotation at zero external magnetic field, solely via
application of appropriate voltages $V_{\mathrm{p}}$ to the piezoelectric
actuator. In the second field range $\unit[0]{mT}\lesssim\mu_{0}H_{\mathrm{meas}}\lesssim\unit[8]{mT}$
in Fig.~\ref{fig:ME-RE}, the angular range within which the magnetization
orientation can be rotated by changing $V_{\mathrm{p}}=\unit[+30]{V}\rightarrow\unit[-30]{V}$
continuously increases, but the magnetization reorientation is not
reversible, since $M_{\mathrm{B}}\left(H\right)\neq M_{\overline{\mathrm{B}}}\left(H\right)$.
This observation can also be consistently understood in a macrospin
model. Here, $\mu_{0}H_{\mathrm{prep}}=\unit[-50]{mT}$ yields $\mathbf{M}_{\mathrm{B}}$
antiparallel to $\mu_{0}H_{\mathrm{meas}}>\unit[0]{mT}$ aligned along
$0^{\circ}$, i.e., $\mathbf{M}_{\mathrm{B}}$ resides in a local
minimum of $F$ at $V_{\mathrm{p}}=\unit[+30]{V}$ and thus in a metastable
state. Sweeping $V_{\mathrm{p}}$ from $\unit[+30]{V}$ to $\unit[-30]{V}$
yields $\mathbf{M}_{\mathrm{C}}$ aligned along the global minimum
of $F$. However, upon increasing the voltage back to $\unit[+30]{V}$,
the magnetization does not rotate back in the same way, but evolves
into the global minimum of $F$ close to $0^{\circ}$ (for details
of the quantitative evolution of $F$ as a function of $V_{\mathrm{p}}$,
see Ref.~\onlinecite{Weiler:NJP:11:2009}). Therefore, the voltage sweep
$V_{\mathrm{p}}=\unit[+30]{V}\rightarrow\unit[-30]{V}\rightarrow\unit[+30]{V}$
results in an irreversible magnetization-orientation change with $\mathbf{M}_{\mathrm{B}}$
and $\mathbf{M}_{\overline{\mathrm{B}}}$ essentially being antiparallel
at $V_{\mathrm{p}}=\unit[+30]{V}$. The third magnetic field range
$\mu_{0}H_{\mathrm{meas}}\gtrsim\unit[10]{mT}$ exceeds $\mu_{0}H_{\mathrm{c}}$
for $V_{\mathrm{p}}=\unit[+30]{V}$, resulting in a (nearly) parallel
alignment of $\mathbf{H}_{\mathrm{meas}}$ and $\mathbf{M}_{\mathrm{B}}$.
Here, the evolution of $M\left(V_{\mathrm{p}}\right)$ is analogous
to the above described for $\mu_{0}H_{\mathrm{meas}}\lesssim\unit[0]{mT}$,
i.e., sweeping $V_{\mathrm{p}}=\unit[+30]{V}\rightarrow\unit[-30]{V}\rightarrow\unit[+30]{V}$
rotates $\mathbf{M}_{\mathrm{B}}$ to $\mathbf{M}_{\mathrm{C}}$ and
back to $\mathbf{M}_{\mathrm{B}}\parallel\mathbf{M}_{\overline{\mathrm{B}}}$.
The angle of rotation decreases with increasing magnetic field strength.

Overall, these findings demonstrate that the macrospin model cannot
only be applied to describe the $M\left(H\right)$ and $R\left(H\right)$
measurements, but also to the $M\left(V_{\mathrm{p}}\right)$ and
$R\left(V_{\mathrm{p}}\right)$ measurements---except for a narrow
range around the coercive field.

\section{Conclusion}

In conclusion, we have studied the applicability and limitations of
a Stoner-Wohlfarth type macrospin model for the description of changes
in the magnetic configuration and magnetoresistance of ferromagnetic/ferroelectric hybrid
systems. To this end, we investigated the magnetic properties in Ni
thin film/piezoelectric actuator hybrids using simultaneous spatially
resolved MOKE and integral magnetotransport measurements at room temperature.
Using dedicated measurement sequences to suppress strain-induced contributions
to the Kerr signal, the imaging of the magnetization state both as
a function of magnetic field and electrical voltage applied to the
piezoelectric actuator becomes possible. We extract an effective magnetization
orientation (macrospin) by spatially averaging the Kerr images. For
experiments both as a function of $H$ and of $V_{\mathrm{p}}$, we
find very good agreement between the AMR calculated using the macrospin
and the measured AMR. Our results show that the magnetization continuously
reorients by coherent rotation---except for $\mathbf{H}$ along a
magnetically easy direction in a very narrow region around the magnetic
coercive field, where the magnetization reorientation dominantly evolves
via domain nucleation and propagation. Taken together, on length scales
much larger than the magnetic domain size, a SW type macrospin model
for both $M\left(H\right)$ and $M\left(V_{\mathrm{p}}\right)$ adequately
describes the corresponding $R\left(H\right)$ and $R\left(V_{\mathrm{p}}\right)$.

\begin{acknowledgments}
Financial support via DFG Project No. GO 944/3-1 and the German Excellence
Initiative via the {}``Nanosystems Initiative Munich (NIM)'' are
gratefully acknowledged.
\end{acknowledgments}


\end{document}